\documentclass[11pt,a4paper]{article}
\pdfoutput=1
\usepackage{jheppub} 
\usepackage{xspace}
\usepackage{subfigure}
\usepackage{multirow}

\newcommand{\ra} {\rightarrow}

\newcommand{\be}{\begin{equation}}
\newcommand{\ee}{\end{equation}}

\newcommand{\bea}{\begin{eqnarray}}
\newcommand{\eea}{\end{eqnarray}}
\newcommand{\beanon}{\begin{eqnarray*}}
\newcommand{\eeanon}{\end{eqnarray*}}
\newcommand{\ba}{\begin{array}}
\newcommand{\ea}{\end{array}}
\newcommand{\bd}{\begin{description}}
\newcommand{\ed}{\end{description}}
\newcommand{\bi}{\begin{itemize}}
\newcommand{\ei}{\end{itemize}}
\newcommand{\ben}{\begin{enumerate}}
\newcommand{\een}{\end{enumerate}}
\newcommand{\bc}{\begin{center}}
\newcommand{\ec}{\end{center}}

\newcommand{\ordEW}{\mathcal{O}(\alpha_{\scriptscriptstyle EM}^6)\xspace}
\newcommand{\ordQCD}{\mathcal{O}(\alpha_{\scriptscriptstyle EM}^4
  \alpha_{\scriptscriptstyle S}^2)\xspace}

\newcommand{\eqn}[1]{Eq.(\ref{#1})}

\newcommand{\eqnsc}[2]{Eqs.(\ref{#1},~\ref{#2})}
\newcommand{\tbn}[1]{Tab.~\ref{#1}}

\newcommand{\fig}[1]{Fig.~\ref{#1}}

\newcommand{\rf}[1]{Ref.~\cite{#1}}
\newcommand{\Phantom}{{\tt PHANTOM}\xspace}

\newcommand{\mc}[1]{\mathcal{#1}}

\newcommand{\ifb}{\mbox{ fb}^{-1}}

\title{
        Exploring alternative symmetry breaking mechanisms at the LHC
with 7, 8 and 10 TeV total energy.
}

\author[a]{Alessandro Ballestrero,}
\author[a,b]{Diogo Buarque Franzosi,}
\author[a,b]{and Ezio Maina}

\affiliation[a]{INFN, Sezione di Torino,\\
Via Giuria 1, 10125 Torino, Italy}
\affiliation[b]{Dipartimento di Fisica, Universit\`a di Torino,\\
Via Giuria 1, 10125 Torino, Italy}

\emailAdd{ballestrero@to.infn.it}
\emailAdd{buarque@to.infn.it}
\emailAdd{maina@to.infn.it}

\abstract{
In view of the annnouncement that in 2012 the LHC  will run at 8 TeV,  we 
study the possibility of detecting signals of alternative mechanisms of ElectroWeak Symmetry Breaking,
described phenomenologically by unitarized models, at energies lower than 14 TeV.
A complete calculation with six fermions
in the final state is performed using the \Phantom event generator.
Our results indicate that at 8 TeV some of the scenarios with TeV scale resonances are likely to be identified
while models with no resonances or with very heavy ones will be inaccessible, unless the available luminosity
will be much higher than expected.
}

\begin{document}

\maketitle
        
\section{Introduction}
\label{sec:intro}

Tantalizing hints of a 125 GeV Higgs boson have been recently reported by both
ATLAS, CMS \cite{ATLAS_Higgs_2011,CMS_Higgs_2011} and, more recently, by CDF and D0 \cite{CDF_D0_HIGGS_2011}.
However the evidence is not yet conclusive
and the possibility that the excess of  events is nothing more than a statistical fluctuation cannot be ruled out.
In the meanwhile the allowed range for the Higgs mass continues to shrink.
In this context,
the role played by high energy vector boson scattering,
either as the final test of the nature of the Higgs boson or
as the main testing ground for Beyond the Standard Model descriptions of ElectroWeak Symmetry Breaking (EWSB),
remains as crucial as ever. 

In previous works \cite{Ballestrero:2009vw,Ballestrero:2010vp,Ballestrero:2011pe}, we have shown that at 14 TeV
the LHC will very probably be able to determine whether the symmetry breaking sector interacts strongly.
If there are heavy resonances around the TeV scale, with $50\ifb$ of integrated luminosity it will be possible to
observe an excess of events in Vector Boson Scattering (VBS)
with respect to the SM predictions.
If no heavy resonances are present or they are much heavier than the accessible scale,
the LHC, with a higher luminosity of about $100\ifb$,
will still produce an excess of events sufficient to determine
the strong nature of the symmetry breaking sector.
If a Higgs is discovered, distinguishing a composite Higgs from an elementary and weakly coupled one
using only VBS data may require a very large luminosity, possibly above $400\ifb$. 

The LHC is scheduled to operate at low energy until the end of 2012. It has been recently announced that
the center of mass energy for the 2012 run will be 8 TeV, as widely expected. 
In this paper we discuss the possibility of detecting signals of unitarized models of  EWSB at the LHC 
with 7, 8 and 10 TeV total energy. The 7 TeV case corresponds to the energy of the 2011 run.
To our knowledge an analysis of VBS based on last year data set has not been published, yet.
Despite the modest luminosity, it would provide a useful warm up exercise and allow validation of the 
theoretical description of the dominant backgrounds.
The 10 TeV case
refers to the possibility that after the long shutdown following the 2012 run, the LHC might resume operation at an
energy lower than 14 TeV. A comparison of the results for the three energies with those at 14 TeV presented in 
\cite{Ballestrero:2011pe} illustrates the effects of the LHC energy on these kind of studies where high mass final states
are looked for.

QCD corrections to boson--boson production via vector boson fusion
\cite{Jager:2006zc,Jager:2006cp,Bozzi:2007ur,Jager:2009xx}
at the LHC have been computed and turn out to be below 10\%. VBFNLO \cite{Arnold:2008rz},
a Monte Carlo program for vector boson fusion, double and triple vector boson production
at NLO QCD accuracy, limited to the leptonic decays of vector bosons, has been
released.
First results for the NLO corrections to $W+4j$ production
have appeared \cite{Berger:2010zx}.

\section{Unitarized Models and their parameters}
\label{sec:unitarized}

As an alternative to full model building it is possible to capture the generic behaviour 
of any symmetry breaking scheme using EffectiveField Theory (EFT) methods, in particular the ElectroWeak Chiral Lagrangian (EWChL)
\cite{Appelquist:1980vg,Longhitano:1980iz,Longhitano:1980tm,Appelquist:1993ka,Contino:2006nn,Giudice:2007fh,Barbieri:2007bh}.
The EWChL provides a systematic
expansion of the full unknown Lagrangian in terms of the fields which are
relevant at energies much lower than the symmetry breaking scale and does not require a detailed
knowledge of the full theory. 

Introducing the matrix
\begin{equation}
\Sigma(x)=\exp\left(\frac{i\sigma^a\omega^a(x)}{v}\right),
\end{equation}
where $\sigma^a$ are the Pauli matrices and $v\approx 246$ GeV is the decay constant of
the Goldstone bosons $\omega^a(x)$ ($a=1,2,3$), which gives the correct masses to the vector ones, 
the only two dimension-4 operators which respect
all required symmetries and are relevant for the study of VBS are:

\begin{align}
\label{eq:ewchlhighorder1}
\mc{L}_4&=\alpha_4\mbox{Tr}[V^\mu,V^\nu]^2 ,\\
\label{eq:ewchlhighorder2}
\mc{L}_5&=\alpha_5\mbox{Tr}[V_\mu,V^\mu]^2 ,
\end{align}
where $V_\mu\equiv (D_\mu\Sigma)\Sigma^\dagger$.

It is then possible to apply Unitarization Methods,
using the lowest order terms in the scattering amplitudes
as building blocks of all order expressions which
respect unitarity and agree up to a finite order with the perturbative result.

A number of unitarization schemes have been implemented in the \Phantom event generator \cite{Ballestrero:2007xq},
within a full six partons in the final state framework, as described in \cite{Ballestrero:2011pe} to which we refer for additional details.

Possible models are characterized by the unitarization scheme and
by the values of the chiral parameters $\alpha_4,\alpha_5$ in \eqnsc{eq:ewchlhighorder1}{eq:ewchlhighorder2}. 
The value of these parameters affect the low energy predictions, and therefore are constrained by data.

\begin{figure}[htb]
\centering
\includegraphics[width=0.6\textwidth]{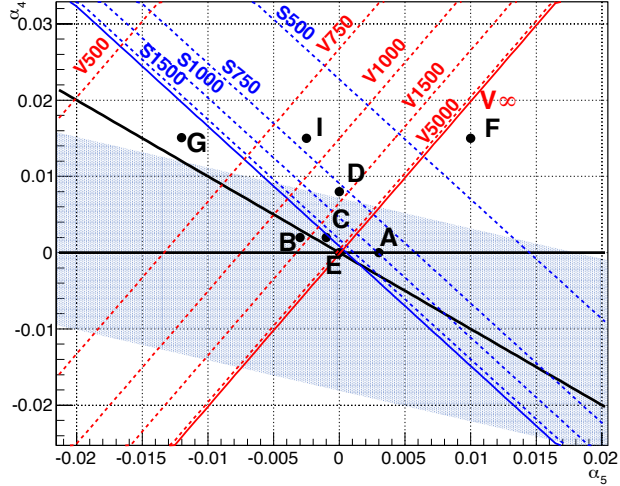}
\caption{Allowed region for the $\alpha_4,\alpha_5$ parameters\cite{Dawson:1994fa,vanderBij:1997ec,Fabbrichesi:2007ad}.
The dotted lines indicate the mass of resonances generated in the IAM method, and the solid blue and red lines
give the limits below which no resonance of the corresponding type is generated. The two solid black lines correspond to the
unitarity and causality constraints derived in \cite{Fabbrichesi:2007ad}.}
\label{fig:alpha_const}
\end{figure}

The most stringent constraints come from their contribution to the $T$-parameter\cite{Peskin:1990zt}. 
In calculations where dimensional regularization has been used \cite{Dawson:1994fa,Brunstein:1996fz}, the logarithmic
divergent contributions to the $T$-parameter are weakly dependent on the cut-off scale and are small, according to \cite{Eboli:2006wa}:
$-0.32<\alpha_4<0.085$ and $-0.81<\alpha_5<0.21$ at $99\%$ CL with a cut-off scale $\Lambda=2$ TeV.
In \cite{vanderBij:1997ec}, it was argued that there are quadratic divergences hidden by the dimensional regularization procedure,
then, using a higher-derivatives regularization, the quadratic divergent contributions to the $T$-parameter have been derived.
The resulting allowed band in the $(\alpha_4 , \alpha_5 )$ plane is depicted in blue in \fig{fig:alpha_const}, where $\Lambda_B=2$ TeV has been assumed, which we intend as a lower limit for this parameter.
Arguments based on unitarity and causality also constrain these parameters
\cite{Fabbrichesi:2007ad} and the associated limits are indicated with the black lines.
In \fig{fig:alpha_const}  we also show the vector and scalar resonance masses produced by the Inverse Amplitude Method as a function of the chiral parameters in the scenarios we have studied in \cite{Ballestrero:2011pe} and here.

\begin{figure}[!htb]
\centering
\includegraphics[width=0.6\textwidth]{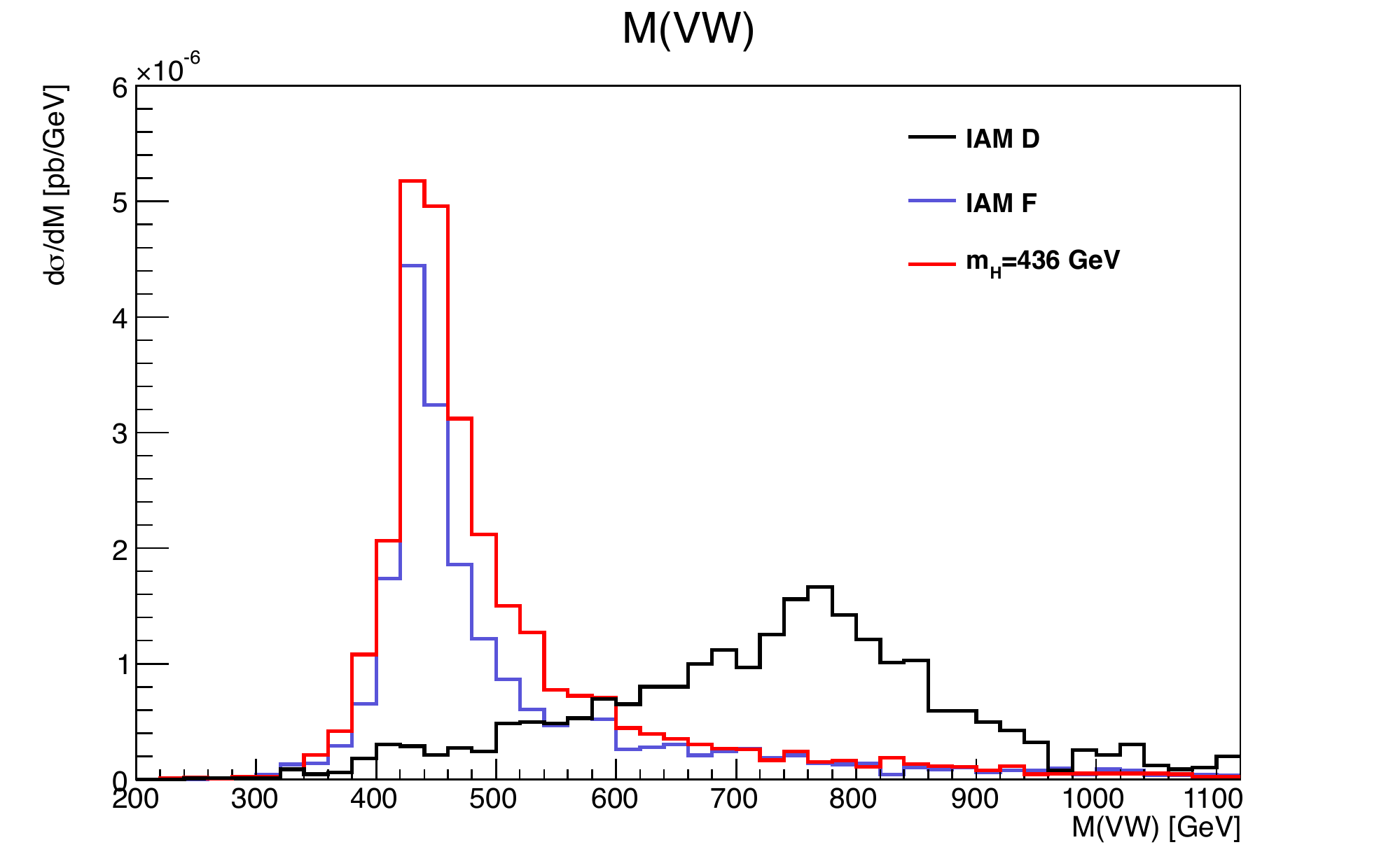}
\caption{$VV$-system mass distribution in the $\ell\nu+4$ jets channel at 7 TeV at $\ordEW$ for the IAM D, IAM F models
and for the SM with $m_H =$ 436 GeV.
The selection cuts shown in \tbn{tbn:cuts} have been applied.}
\label{fig:MVW_DFH}
\end{figure}

For our present study,
we have selected three representative scenarios of a strong symmetry breaking sector, all of them are instances of
the Inverse Amplitude Method (IAM) of unitarization.
The IAM procedure has the nice feature that, beside the perturbative expansion of the amplitude obtained from the EWChL plus the additional terms in \eqn{eq:ewchlhighorder1}, no additional parameters is used.
The resonances are produced by the method itself. 
On the contrary, in both the KM and N/D method resonant states can be arbitrarily introduced by hand.
We have considered one scenario without resonances, with chiral parameters $(0,0)$, called IAM E,
which is slightly enhanced with respect to the no-Higgs scenario due to the higher order terms. This model has the smallest cross section for the processes we study here, while in the $2j\ell^\pm\ell^\pm\nu\nu$ channel, where no resonance is present, its production rate is typically larger than the rate in models which contain resonant states
\cite{Ballestrero:2011pe}.
As a consequence, the IAM E model is the most difficult one to detect  among all instances we have examined for this study and represents the lower limit for the possible effects of unitarized models. 
In the second scenario, IAM G with parameters $(15,10)\times 10^{-3}$, a 0.6 TeV vector resonance dominates the scattering cross section. This is an example of models with light resonances.
A third scenario, IAM J with parameters $(9,-3)\times 10^{-3}$,
contains a scalar and a vector resonance, both  at 1 TeV, and can be taken as representative of models with relatively heavy resonances which, however, are not so heavy to be totally undetectable at the energies which will be available next year.

Models in which only scalar resonances are expected have been neglected  because their predictions are 
very similar to those obtained in the SM for a Higgs of the same mass and therefore,
the prospect for their discovery can be inferred from the detailed studies dedicated to the SM Higgs searches.
This behaviour is demonstrated in \fig{fig:MVW_DFH} where the mass distribution of the $VV$-system 
in the $\ell\nu+4j$  channel is shown for three different scenarios at 7 TeV.
The IAM F scenario, with parameters $(15,10)\times 10^{-3}$,
admits a 436 GeV scalar resonance. It is compared with
the SM with a 436 GeV Higgs boson and
with the IAM D scenario, $(8,0)\times 10^{-3}$, studied in a previous work \cite{Ballestrero:2011pe},
with a scalar resonance at 0.8 TeV and a vector one at about 1.4 TeV.
Comparing \fig{fig:MVW_DFH}  with Fig. 2 of \rf{Ballestrero:2011pe}, it can be noticed that the
1.4 TeV resonance has essentially disappeared. As a consequence we have limited ourselves to models
with relatively light resonances.

\section{Results}
\label{sec:results}

We have concentrated on the three final states which are most relevant for detecting strong scattering signals:
the $\ell\nu+4$ jets semi-leptonic channel, the $2jW^+W^-\ra 2j\ell^+\ell^-\nu\bar{\nu}$ channel and
the $3\ell\nu+2$ jets channel which is useful in the search for vector resonances.
In \tbn{tbn:cuts}, we show the set of kinematical cuts applied in each of these channels in order to enhance the
discrepancy between the strong scenarios and the predictions of the Standard Model with a light Higgs and to improve the signal to background ratio. In this paper we have taken the Higgs mass to be 170 GeV. The exact value of this parameter,
within the limits derived from precision data, is immaterial.  
A detailed analysis of kinematical cuts and of the optimization of exclusion probabilities is presented in
\cite{Ballestrero:2009vw,Ballestrero:2010vp,Ballestrero:2011pe}.
Here we use a looser set of cuts to compensate for the lower energies. It should be pointed out that,
while a cut based treatment is perfectly adequate for a preliminary analysis at parton level, 
a more sophisticated Multi Variate Analysis would certainly provide better results. On the other hand a
more realistic implementation of experimental uncertainties and hadronization effects
would in all likelihood work in the opposite direction.
The results of all channels, including those not discussed in the following, could be combined in order to improve the
sensitivity.

\begin{table}[tbh!]
\centering
\begin{tabular}{|l|l|}
\hline
$p_T(j)>30$ GeV 		& $p_T(\ell)>70/70/20$ GeV 	\\
\hline
$p_T^{miss}>70/20/20$ GeV 	& $p_T(j_c)>70$ GeV  \\
\hline
 $\eta(j)<6.5$			& $\eta(\ell)<2/2/3$  \\
\hline
$\Delta\eta(j_fj_b)>4/4/3$ 	& $\Delta\eta(V_{rec}j)>0.6$ \\
\hline
$\Delta R(jj)>0.3$ 		& $M(\ell\ell)>20$ GeV \\
\hline
$M(jj)>60$ GeV			& $M(j_fj_b)>700/600/100 $ GeV\\
\hline
$p_T(V_{rec})>70/100$ GeV	& $|M(V_{rec}j)-M_{TOP}|>15$ GeV \\
\hline
\hline
$M(j\ell)>180$ GeV 		& $|p_T(\ell^+)-p_T(\ell^-)|>100$ GeV \\
\hline
\end{tabular}
\caption{Kinematical cuts applied on the analysis. Different values correspond to different channels in the order
$4j\ell\nu$, $2j\ell\ell\nu\nu$ and $2j3\ell\nu$. 
$j_f,j_b$ refer to the most forward and most backward of the jets.
$j_c$ indicates one of the central jets in the $4j\ell\nu$ channel. 
$V_{rec}$ stands for the boson which is reconstructed from the lepton and neutrino momenta,
the latter obtained from the requirement that $(p_\ell + p_\nu )^2 = M_W^2$
and is meaningful only for $4j\ell\nu$ and $2j3\ell\nu$. 
The constraints on the last line apply only to the $2j\ell\ell\nu\nu$ channel.
}
\label{tbn:cuts}
\end{table}

%%%%%%%%%%%%%%%%%%%%%%%%%%%%%%%%%%%%%%%%%%%%%%%%
%% lv + 4 jets
%%%%%%%%%%%%%%%%%%%%%%%%%%%%%%%%%%%%%%%%%%%%%%%%

\begin{figure}[b!ht]
\centering
\includegraphics[width=\textwidth,height=5cm]{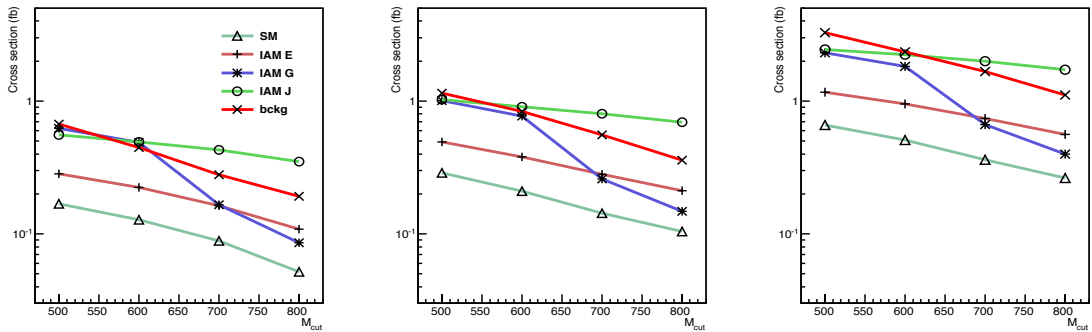}
\caption{Cross section in femtobarns after all selection cuts  in \tbn{tbn:cuts}
for 7, 8 and 10 TeV of center of mass energy 
at four different values of
the minumum invariant mass of the reconstructed $VV$-system in the $\ell\nu+4$ jets channel
for the IAM E, IAM G and IAMJ models. The lines are only meant to guide the eye.
For comparison we show also the SM predictions for $m_H =$ 170 GeV and the background
from $V+4j$ and $t\overline{t}+jets$ production.}
\label{fig:xsec_lv4j}
\end{figure}

The semi-leptonic channel with one charged lepton, electron or muon, and four jets in the final state gives the best
discriminating power because of its high rates.
Both vector and scalar resonances are produced. \fig{fig:xsec_lv4j} shows the
cross section after all selection cuts in \tbn{tbn:cuts}, for 7, 8 and 10 TeV  center of mass energy and
for different minumum invariant mass of the
$VV$ reconstructed system, whose distribution for $\sqrt{s}$ = 10 TeV is shown on the left side of \fig{fig:4jlv_mvv}.
At 7 TeV, with $25\ifb$ and $M_{min}(VV) >$ 500 GeV, 7 events above background can be expected
for the IAM E model, to be compared with a SM EW prediction of 4 events and a background of about 16 events.
These numbers increase to 29 and 16 at 10 TeV for the IAM E model and for the SM respectively.
The corresponding predictions for the background is 86 events.
The IAM E scenario is the most unfavourable one since it does not predict any type of resonance. 

The Probability Distribution Functions (PDFs) of the number of events expected for each scenario at $50\ifb$ of integrated luminosity, with
10 TeV of center of mass energy and considering $VV$-system masses above 500 GeV is shown on the right side of \fig{fig:4jlv_mvv}.
Here and in the following, the PDF's are computed assuming  Poissonian statistical 
fluctuations of the number of events computed by the MC and a theoretical
error on the number of signal events which we model as a flat distribution of
$\pm 30\%$ from the actual value  \cite{Ballestrero:2009vw,Ballestrero:2010vp}.
It is important to remark that we consider $W+4$ jets and $t\bar{t}+$2jets as backgrounds which can be precisely measured in complementary regions of phase space,
hence unnaffected by theoretical uncertainties. The contributions from $\ordEW+\ordQCD$, on the other hand,
which describe the production of two vector bosons in association with a pair of jets,
are affected by both theoretical and statistical uncertainties.

In the plot the vertical line represents the 95\% limit of the
light Higgs distribution. We therefore compute what we call the PBSM@95\%CL
(Probability Beyond the SM at 95\% Confidence Level)
for the various scenarios as the probability that a number of events larger
than the 95\% limit occurs. 

The PBSM@95\%CL assuming one of
the three alternative models is reported in \tbn{tbn:4jlv_pbsm}.
The IAM G and IAM J scenarios, with resonances at or below one  TeV,  have a better than 70\% chance to yield results
outside the 95\% CL for the SM already at 7 TeV with 25 $\ifb$, which grows larger than 90\% at 10 TeV.
Increasing the energy from 7 to 8 TeV has a modest effect for the IAM G scenario while it is more beneficial
for the IAM J case with its heavier resonances. The IAM E scenario, and all cases with no resonances or very heavy ones,
requires higher luminosities: about 50  $\ifb$ at 10 TeV and about 200  $\ifb$ at 8 TeV to reach a probability of at least 50\% 
to exceed the SM 95\% CL.

\begin{figure}[tbh!]
\centering
\includegraphics*[width=0.49\textwidth,height=5cm]{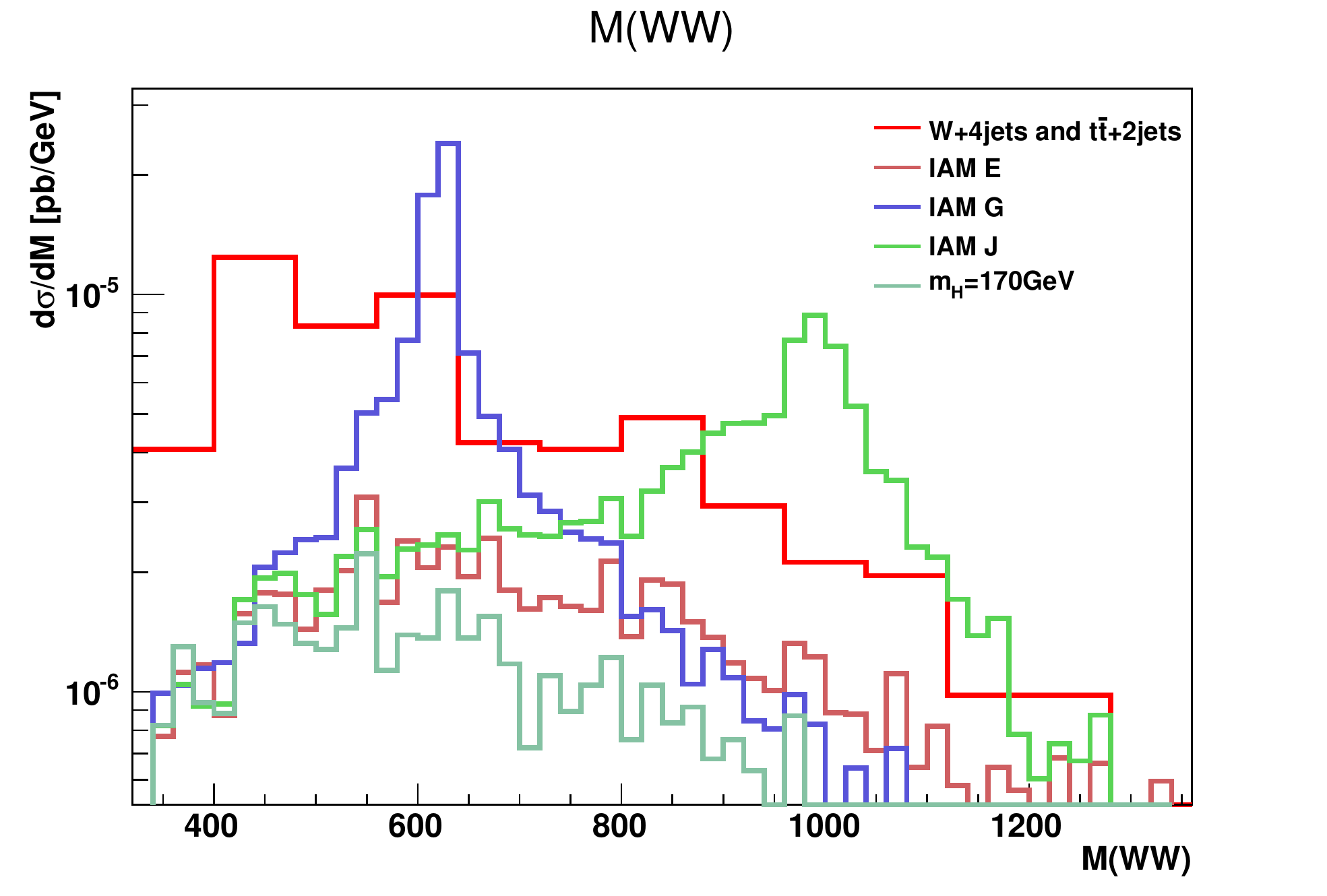}
\includegraphics*[width=0.49\textwidth,height=5cm]{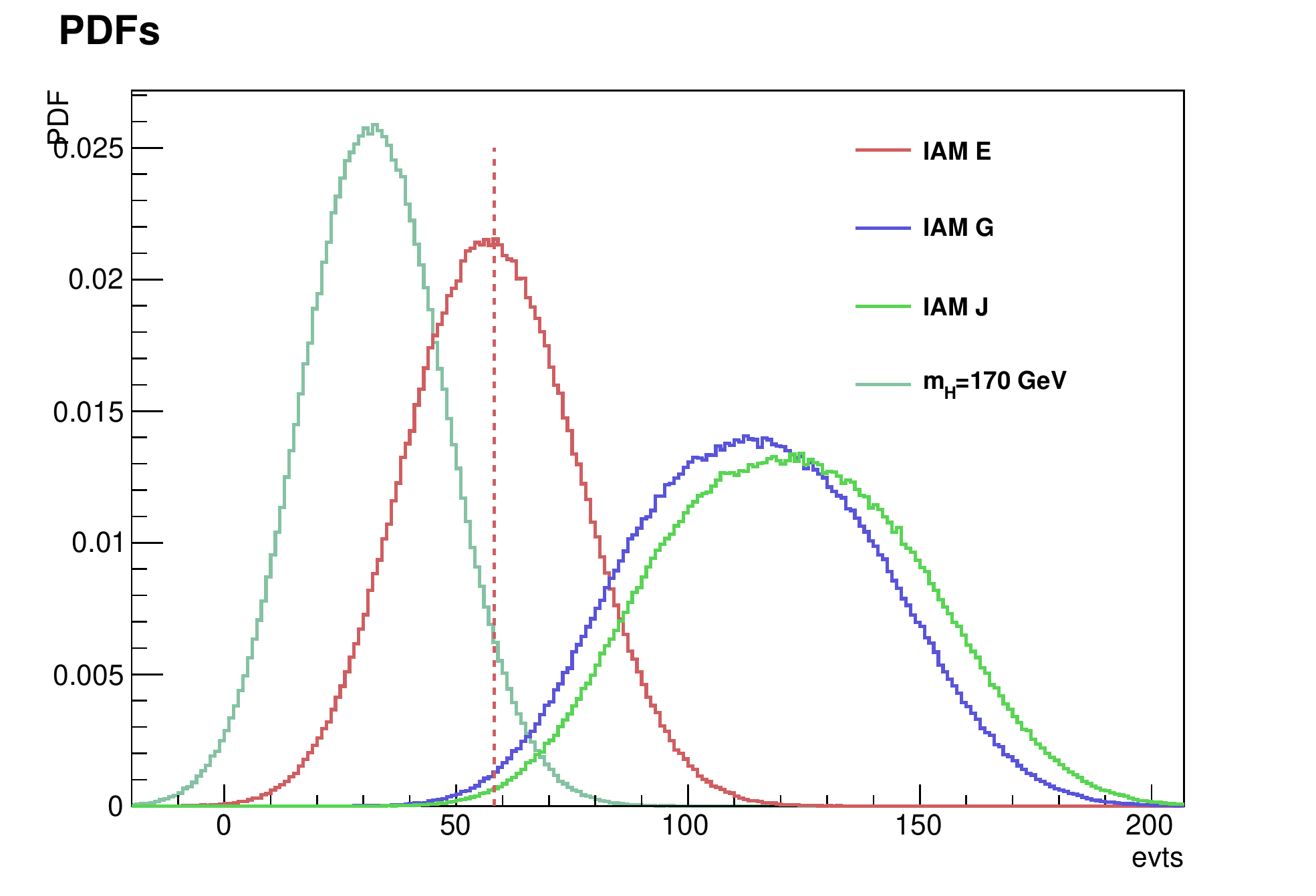}
\caption{On the \emph{left}, the invariant mass distribution of the two central jets,
        the lepton and the reconstructed neutrino (according to prescription of \cite{Ballestrero:2009vw}) $M(j_cj_c\ell\nu)$
        for $\sqrt{s}$ = 10 TeV.
On the \emph{right}, probability distribution of the number of events above the measured background
for $50\ifb$ and $M_{min}(VV) >$ 500 GeV. The vertical line indicates the 95\%CL in the SM.}
\label{fig:4jlv_mvv}
\end{figure}

%\begin{table}[tbh!]
%\centering
%\begin{tabular}{|c|c|c|c|c|c|c|c|c|c|c|}
%\hline%
%	& \multicolumn{2}{|c|}{$m_H=170$ GeV}& \multicolumn{2}{|c|}{IAM E} & \multicolumn{2}{|c|}{IAM G} & \multicolumn{2}{|c|}{IAM J} & \multicolumn{2}{|c|}{backg} \\	
%\hline
%M\textbackslash E& 7 	& 10    & 7     & 10    & 7    & 10     & 7     & 10    & 7    & 10  \\
%\hline
%500	& .169  & .658	&.283   & 1.17	& .625 & 2.31	& .557	& 2.45	& .629 & 3.44 \\
%\hline
%600	&.128   & .508	& .223  & .948	& .489	& 1.82	& .494	& 2.24	&.435 & 2.27 \\
%\hline
%700	&.0888  & .363	&.163   & .737	& .166	& .664	& .428	& 1.99	&.321 & 1.76 \\
%\hline
%800	& .0518 & .264	&.109	& .562	& .086	& .398 	& .349  & 1.72	&.235 & 1.29 \\
%\hline
%\end{tabular}
%\caption{Cross sections in femto-barns for $\ell\nu+4j$ with different cuts on the invariant mass of $VV$ reconstructed system (M) and for 
%center of mass energy, E=7 TeV and 10 TeV. 
%The contributions for each scenario correspond to the sum of $\ordEW+\ordQCD$ and backg represents the sum 
%of $t\bar{t}+$2jets and $W+$4jets. }
%\label{tbn:4jlv_xsection}
%\end{table}

\begin{table}[tbh!]
\centering
\begin{tabular}{|c|c|c|c|c|c|c|c|c|c|c|c|}
\hline
	& \multicolumn{3}{|c|}{IAM E} & \multicolumn{3}{|c|}{IAM G} & \multicolumn{3}{|c|}{IAM J}	\\
\hline
L\textbackslash E	
        & 7     &  8    & 10      & 7  &  8    & 10 & 7 & 8 & 10  \\
%\hline
%$M_{cut}$& 500  & 600   & 500   & 500   & 800	& 800 \\		
\hline
25	& 16.06 $^a$ & 19.03 $^a$ & 35.37 $^b$ & 71.10 $^a$ & 75.48 $^a$ & 93.80 $^a$ & 73.32 $^d$ & 81.77 $^d$ & 99.32 $^d$  \\
\hline
50	& 22.70 $^a$ & 27.88 $^a$ & 51.56 $^b$ & 89.14 $^a$ & 91.68 $^a$ & 99.12 $^a$ & 91.55 $^d$ & 95.62 $^d$ & 99.99 $^d$  \\
\hline
100	& 33.51 $^a$ & 41.08 $^a$ & 69.28 $^c$ & 97.85 $^a$ & 98.54 $^a$ & 99.97 $^a$ & 98.89 $^d$ & 99.66 $^e$ & 100 $^d$ \\
\hline
200	& 48.25 $^b$ & 57.08 $^a$ & 83.44 $^c$ & 99.87 $^a$ & 99.93 $^a$ & 100 $^a$   & 99.97 $^d$ & 100 $^d$   & 100 $^d$ \\
\hline
\end{tabular}
\caption{PBSM@95\%CL in the $\ell\nu+4$ jets channel
with 25, 50, 100 and 200 $\ifb$ of integrated luminosity, L.
For each luminosity and model we have used the mass cut which gives the best probability.
They are specified by the superscript according to the following scheme:
$^a$, $^b$, $^c$, $^d$, $^e$ for 500, 600, 700, 800, 900 GeV respectively}
\label{tbn:4jlv_pbsm}
\end{table}

%\begin{table}[tbh!]
%\centering
%\begin{tabular}{|c|c|c|c|c|c|c|c|c|}
%\hline%
%	& \multicolumn{2}{|c|}{IAM E} & \multicolumn{2}{|c|}{IAM G} & \multicolumn{2}{|c|}{IAM J}	\\
%\hline
%L\textbackslash E	
%        & 7         & 10      & 7     & 10& 7& 10  \\
%%\hline
%$M_{cut}$& 500  & 600   & 500   & 500   & 800	& 800 \\		
%\hline
%25	& 16.06 (0.5) & 35.37 (0.6) & 71.10 (0.5) & 93.80 (0.5) & 73.32 (0.8) & 99.32 (0.8)  \\
%\hline
%50	& 22.70 (0.5) & 51.56 (0.6) & 89.14 (0.5) & 99.12 (0.5) & 91.55 (0.8) & 99.99 (0.8)  \\
%\hline
%100	& 33.51 (0.5)  & 69.28 (0.7) & 97.85 (0.5) & 99.97 (0.5) & 98.89 (0.8) & 100 (0.8) \\
%\hline
%200	& 48.25 (0.6) & 83.44 (0.7) & 99.87 (0.5) & 100 (0.5)  & 99.97 (0.8) & 100 (0.8) \\
%\hline
%\end{tabular}
%\caption{Probability in percent to exclude the standard at 95\% of confidence level with 25, 50, 100 and 200 $\ifb$ of integrated luminosity, L.
%In each case we have used the most favorable mass cut which is given inside parentheses in TeV.}
%\label{tbn:4jlv_pbsm_2}
%\end{table}

%%%%%%%%%%%%%%%%%%%%%%%%%%%%%%%%%%%%%%%%%%%%%%%%
%% WW + 2 jets -> 2jlvlv
%%%%%%%%%%%%%%%%%%%%%%%%%%%%%%%%%%%%%%%%%%%%%%%%

\begin{figure}[!htb]
\centering
\includegraphics[width=\textwidth,height=5cm]{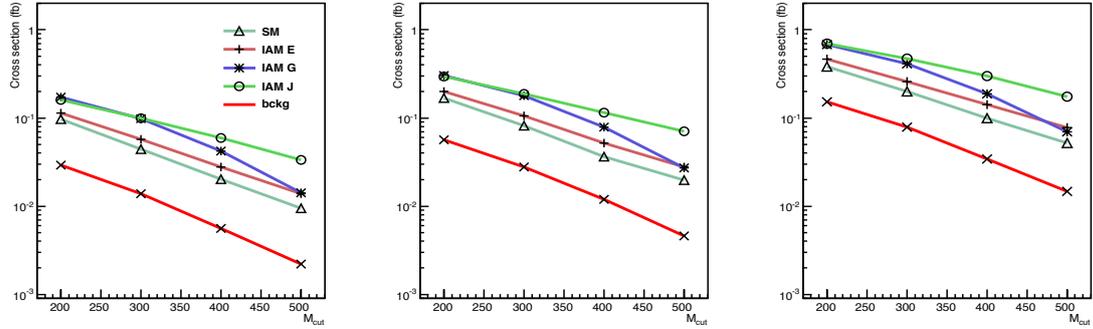}
\caption{Cross section in femtobarns after all selection cuts for 7, 8 and 10 TeV of center of mass energy
at four different values of
        the minumum invariant mass of the lepton pair in the $(WW)\ell\nu\ell\nu+2j$ channel
        for the IAM E, IAM G and IAMJ models. For comparison we show also the SM predictions for $m_H =$ 170 GeV and the background from $t\overline{t}+jets$ production.}
\label{fig:xsec_2l2v2j}
\end{figure}

The leptonic channel in which two $W$ bosons decay to opposite sign leptons, each either  an electron or a muon,
is sensitive  to both scalar and vector resonances 
and is very important in general for the study of strong $WW$ scattering even though 
the invariant mass of the boson pair cannot be directly measured.  
The mass of the charged leptons pair has been shown in \rf{Ballestrero:2011pe} to be an effective variable for the separation of 
unitarized models from the SM.
At 7 TeV  only a handful of events are expected for $L=100\ifb$ of integrated luminosity,
whereas for 10 TeV, with $L=25\ifb$,  12, 10, 6 and 5 events could be produced
for IAM J, IAM G, IAM E and the SM respectively for $M_{\ell\ell^\prime} >$  300 GeV
as can be extracted from the cross sections in \fig{fig:xsec_2l2v2j}.
The corresponding PBSM@95\%CL are reported in \tbn{tbn:ww_pbsm}. Despite the smaller background
this channel is less efficient than the $\ell\nu+4$ jets one discussed previously.
The IAM E model has a probability of less than 15\% of producing an excess even at 10 TeV with a luminosity of 25 $\ifb$.
For the IAM G and IAM J models the probability is below 30\% at 7 TeV and 25 $\ifb$. With this luminosity, which
is slightly more optimistic than the 15 $\ifb$ officially expected in 2012, the probability exceeds 50\% only at 10 TeV.
As obvious, at higher luminosity our predictions are much more optimistic.

%\begin{table}[tbh!]
%\centering
%\begin{tabular}{|c|c|c|c|c|c|c|c|c|c|c|}
%\hline%
%	& \multicolumn{2}{|c|}{$m_H=170$ GeV}& \multicolumn{2}{|c|}{IAM E} & \multicolumn{2}{|c|}{IAM G} & \multicolumn{2}{|c|}{IAM J} & \multicolumn{2}{|c|}{backg} \\	
%\hline
%M\textbackslash E
%	& 7     & 10    & 7     & 10    & 7     & 10    & 7    & 10     &  7   & 10 \\
%\hline
%300	& .044 & .199	& .057 & .260	& .098  & .412	& .099 & .473	& .025 & .076 \\
%\hline
%400	& .020 & .099	& .028 & .141	& .042	& .188	& .060 & .300	& .011 & .044 \\
%\hline
%\end{tabular}
%\caption{Cross sections in femto-barns for $(WW)\ell\nu\ell\nu+2j$ with different cuts on the
%invariant mass of $\ell\ell$-system (M) and for 
%center of mass energy, E=7 TeV and 10 TeV.
%The contributions for each scenario correspond to the sum of $\ordEW+\ordQCD$ and backg represents $t\bar{t}+$2jets. }
%\label{tbn:ww_xsection}
%\end{table}

\begin{table}[tbh!]
\centering
\begin{tabular}{|c|c|c|c|c|c|c|c|c|c|}
\hline
	& \multicolumn{3}{|c|}{IAM E} & \multicolumn{3}{|c|}{IAM G} & \multicolumn{3}{|c|}{IAM J}	\\
\hline
L\textbackslash E	
        & 7          &  8        & 10          & 7           &  8         & 10         & 7          &  8         & 10        \\
\hline
25	& 8.47 $^a$  & 10.44 $^b$ & 14.23 $^a$ & 24.41 $^a$  & 36.49 $^a$ & 51.83 $^a$ & 27.17 $^b$ & 42.69 $^b$ & 65.68 $^a$ 	 \\
\hline
50	& 10.02 $^a$ & 13.06 $^b$ & 18.94 $^a$ & 35.83 $^a$  & 53.23 $^a$ & 70.07 $^a$ & 37.90 $^b$ & 61.95 $^b$ & 84.73 $^b$  \\
\hline
100	& 12.63 $^a$ & 17.34 $^b$ & 26.37 $^b$ & 52.81 $^a$  & 72.07 $^a$ & 84.59 $^a$ & 56.76 $^b$ & 81.94 $^b$ & 95.97 $^b$  \\
\hline
200	& 16.49 $^a$ & 24.08 $^b$ & 36.35 $^b$ & 71.87 $^a$  & 86.74 $^a$ & 93.23 $^a$ & 76.92 $^b$ & 94.91 $^b$ & 99.50 $^b$ \\
\hline
\end{tabular}
\caption{PBSM@95\%CL in the $(WW)\ell\nu\ell\nu+2j$ channel
with 25, 50, 100 and 200 $\ifb$ of integrated luminosity, L.
For each luminosity and model we have used the mass cut which gives the best probability.
They are specified by the superscript according to the following scheme:
$^a$, $^b$ for 300, 400 GeV respectively.}
\label{tbn:ww_pbsm}
\end{table}

%%%%%%%%%%%%%%%%%%%%%%%%%%%%%%%%%%%%%%%%%%%%%%%%
%% 2j3lv
%%%%%%%%%%%%%%%%%%%%%%%%%%%%%%%%%%%%%%%%%%%%%%%%

\begin{figure}[!htb]
\centering
\includegraphics[width=\textwidth,height=5cm]{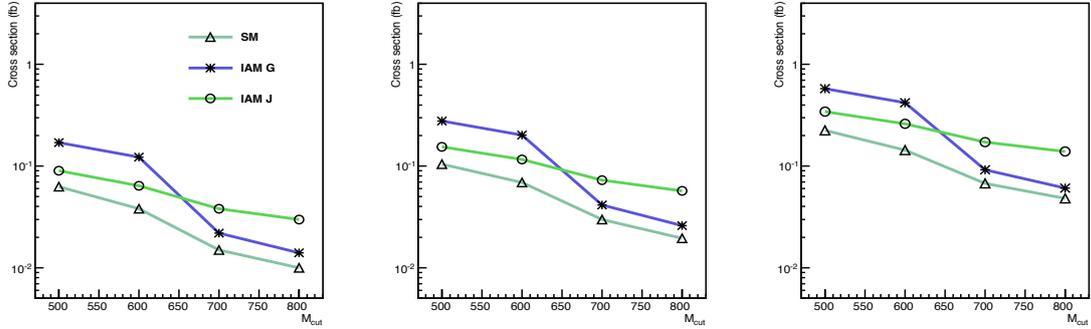}
\caption{
Cross section in femtobarns after all selection cuts for 7, 8 and 10 TeV of center of mass energy
at four different values of
the minumum invariant mass of the reconstructed $VV$-system in the $3\ell\nu+2$ jets channel
for the IAM E, IAM G and IAMJ models. For comparison we show also the SM predictions for $m_H =$ 170 GeV.
}
\label{fig:xsec_3lv2j}
\end{figure}

The three leptons channel can contribute in scenarios in which vector resonances are present,
as is the case of the IAM G and IAM J models.
The cross sections are reported in \fig{fig:xsec_3lv2j} as a function of the minimum reconstructed invariant mass
of the $WZ$ pair.  The corresponding exclusion probabilities are shown in \tbn{tbn:3l_pbsm} for luminosities ranging from
25 to 200 $\ifb$ .
The probability to observe strong scattering in the IAM G case is above 50\% already at 7 TeV with 25 $\ifb$
and grows significantly with the collider energy.
In the IAM J scenario, since the scalar resonance does not appear in this final state, the probability of observing an excess
is markedly smaller.
The probability for the IAM E non-resonant scenario is negligible and has been omitted.

%\begin{table}[tbh!]
%\centering
%\begin{tabular}{|c|c|c|c|c|c|c|}
%\hline
%	& \multicolumn{2}{|c|}{$m_H=170$ GeV} & \multicolumn{2}{|c|}{IAM G} & \multicolumn{2}{|c|}{IAM J}	     \\      
%\hline
%M\textbackslash E
%	& 7     & 10    & 7     & 10     & 7     & 10  \\
%\hline
%500	& .063 & .224	 & .171 & .579	 & .090  & .344	 \\
%\hline
%600	& .038 & .144	 & .122 & .421	 & .064  & .260	 \\
%\hline
%800	& .015 & .068	 & .022	& .092	 & .038 & .172	 \\
%\hline
%900	& .010 & .048	 & .014	& .061	 & .030 & .139	 \\
%\hline
%\end{tabular}
%\caption{Cross sections in femto-barns for $3\ell\nu+2j$ with different cuts on the invariant mass of
%the reconstructed $VV$-system. The contributions for each scenario correspond to the sum of $\ordEW+\ordQCD$. }
%\label{tbn:3l_xsection}
%\end{table}

\begin{table}[tbh!]
\centering
\begin{tabular}{|c|c|c|c|c|c|c|c|c|c|}
\hline
	 & \multicolumn{3}{|c|}{IAM G}	 & \multicolumn{3}{|c|}{IAM J} \\
\hline
L\textbackslash E
 	 & 7          &  8        & 10          & 7           &  8         & 10    \\   
\hline
25	 & 50.53 $^a$ & 63.18 $^a$ & 82.05 $^a$ & 21.74 $^d$ & 29.28 $^d$ & 48.50 $^e$   \\
\hline
50	 & 71.93 $^a$ & 82.12 $^a$ & 93.99 $^a$ & 31.02 $^d$ & 43.24 $^e$ & 68.63 $^e$ \\
\hline
100	 & 88.13 $^a$ & 94.08 $^a$ & 98.97 $^b$ & 43.83 $^e$ & 63.71 $^e$ & 86.42 $^e$ \\
\hline
200	 & 97.09 $^b$ & 98.94 $^b $& 99.95 $^b$ & 63.63 $^e$ & 82.62 $^e$ & 96.56 $^e$\\
\hline
\end{tabular}
\caption{PBSM@95\%CL in the $3\ell\nu+2$ jets channel
with 25, 50, 100 and 200 $\ifb$ of integrated luminosity, L.
For each luminosity and model we have used the mass cut which gives the best probability.
They are specified by the superscript according to the following scheme:
$^a$, $^b$, $^c$, $^d$, $^e$ for 500, 600, 700, 800, 900 GeV respectively}
\label{tbn:3l_pbsm}
\end{table}

\section{Conclusions}

In this paper we have studied the probability of finding a number of events exceeding the 95\%
confidence limit for the Standard Model in unitarized models of Electoweak Symmetry Breaking.
We have focused on models based on the Inverse Amplitude Method.
Our results indicate that at 8 TeV, the energy of the 2012 LHC run, some of the
scenarios with TeV scale resonances are likely to be identified while models with no resonances or with
very heavy ones will be inaccessible. 
In the absence of a positive result, it will be possible to obtain more stringent limits on the values of
$\alpha_4$ and $\alpha_5$, at least whithin a specified unitarization scheme.
If reaching the design energy of 14 TeV will prove more difficult than expected,
an energy of 10 TeV would already significantly increase the LHC reach.
 
\section *{Acknowledgments}

A.B. wishes to thank the Dep. of Theoretical Physics of Torino University
for support. \\
This work has been supported by MIUR under contract 2008H8F9RA\_002 . 
%\hfill*
%\eject
%\newpage

%%%%%%%%%%%%%%%%%%%%%%%%%%%%%%%%%%%%%%%%%%%%%%%%%%%%%%%%%%%%%%%%%%%%%%%%%

% To include bibliography do:
% 1- pdflatex FileName.tex
% 2- bibtex FileName
% 3- pdflatex FileName.tex
% 4- pdflatex FileName.tex
%\bibliographystyle{unsrt}
%\bibliographystyle{hunsrt}
%\bibliographystyle{h-elsevier.bst}
\bibliographystyle{JHEP}
\bibliography{ubib}

\end{document}